\documentclass[12pt]{article} 

\usepackage{scicite}

\usepackage{times}

\usepackage{amsmath}
\usepackage{amsfonts}
\usepackage{amssymb}
\usepackage{graphicx}

\usepackage{booktabs}

\newenvironment{sciabstract}{%
\begin{quote} \bf}
{\end{quote}}

\usepackage[labelfont=bf,figurename=Fig.,labelsep=period,font=footnotesize]{caption}

\title{\Large\bf On the inadequacy of nominal assortativity\\ for assessing homophily in networks}

\author
{
Fariba Karimi${}^{1,\ast}$ and Marcos Oliveira${}^{2,\dagger}$\vspace{.1in}\\
\small{${}^{1}$Complexity Science Hub Vienna, 1080 Vienna, Austria}\\
\small{${}^{2}$Computer Science, University of Exeter, Exeter, United Kingdom}\vspace{.1in}\\
\footnotesize{$^\ast$karimi@csh.ac.at $^\dagger$moliveira@tuta.io}\vspace{-.4cm}
}

\date{}

\usepackage{changepage}

\usepackage{comment}

\usepackage{xcolor}
\usepackage{soul}
\definecolor{reddish}{HTML}{FBB4AE}
\definecolor{blueish}{HTML}{B3CDE3}
\definecolor{magentish}{HTML}{FF00AA}
\definecolor{greenish}{HTML}{a1d99b}

\usepackage[hidelinks]{hyperref}

\usepackage{setspace}

\usepackage{amsmath}           

\usepackage{lineno}

\begin{document} 



\maketitle 





\begin{sciabstract}
Abstract

\textnormal{Nominal assortativity (or discrete assortativity) is widely used to characterize group mixing patterns and homophily in networks, enabling researchers to analyze how groups interact with one another. Here we demonstrate that the measure presents severe shortcomings when applied to networks with unequal group sizes and asymmetric mixing. We characterize these shortcomings analytically and use synthetic and empirical networks to show that nominal assortativity fails to account for group imbalance and asymmetric group interactions, thereby producing an inaccurate characterization of mixing patterns. We propose adjusted nominal assortativity and show that this adjustment recovers the expected assortativity in networks with various level of mixing. Furthermore, we propose an analytical method to assess asymmetric mixing by estimating the tendency of inter- and intra-group connectivities. Finally, we discuss how this approach enables uncovering hidden mixing patterns in real-world networks. }
\end{sciabstract}

\baselineskip24pt 
\section*{Introduction}
Understanding how groups interact in networks is fundamental for uncovering mechanisms underlying diverse phenomena, from protein interactions to social communication\cite{McPherson2001,goodreau2009birds,oliveira2022group}. Such group-level interactions often generate mixing patterns in networks, commonly assessed with single-valued measures such as nominal assortativity\cite{Newman2003mixing,peel2018multiscale}. Though these measures help analyze group mixing concisely, they may be grounded on unrealistic assumptions about the network structure, which might produce imprecise estimates of group mixing tendencies, limiting our understanding of groups in networks. 

Recent advances in relational data collection have enabled studies on mixing patterns to investigate fundamental processes that drive such tendencies\cite{schaible2021sensing}. In particular, much research has characterized homophily---nodes' tendency to connect with alike---in a variety of social settings due to processes such as selective mixing and in-group favoritism\cite{goodreau2009birds,efferson2008coevolution}. For instance, sexual partnership networks are assortative by race in the United States, meaning that individuals form ties with partners of the same race more often than one would expect by chance\cite{Newman2003mixing}. Similarly, college students are more likely to have friendships with peers of the same gender, major, residence, and year\cite{Traud2012}. These homophilic tendencies have been documented in various other social phenomena such as research collaboration\cite{Pepe2010}, artist partnerships\cite{Jacobson2009}, lawmaking\cite{Oliveira2014}, and book readership\cite{Bucur2019}. Beyond social networks, previous works have also demonstrated homophily in biological domains such as networks of protein similarity\cite{cells_network} and dolphin companionship\cite{Lusseau2004}. 

Researchers in network science generally use the so-called nominal assortativity to characterize mixing patterns regarding categorical attributes (e.g., race, gender, protein type)\cite{Newman2003mixing}, such as those mentioned above. Nominal assortativity describes how intra- and inter-group connectivity diverges from what we expect solely due to degree connectivity of the nodes and groups. The advantage of this measure compared to other existing measures of homophily is that it takes into consideration connectivity patterns of groups to assess the statistical significance. Its straightforward definition produces an intuitive quantity that ranges from $-1$ (i.e., complete disassortative mixing) to $0$ (i.e., neutral) to $+1$  (i.e., complete assortative mixing), which enables researchers to analyze group mixing in networks concisely. Recently, Cinelli and colleagues showed that the assortativity coefficient, $r$, is bounded, analogous to the constraints that exist on Pearson correlation coefficients\cite{cinelli2020network}. They demonstrated that the assortativity coefficient can range between $r_{\text{min}}$ and $r_{\text{max}}$, which are dependent on the edge counts in the networks. This specifically implies that $r=1$ is only achievable when the sum of the inter-group edge counts, is equal to the total number of edges in the network. Conversely, $r=-1$ is attainable solely in cases where this sum is zero. Crucially, this work exposes the influence of certain network attributes such as metadata assignment and degree sequence on these bounds of $r$.

\begin{figure}[b!]
\centering
\includegraphics[width=6in]{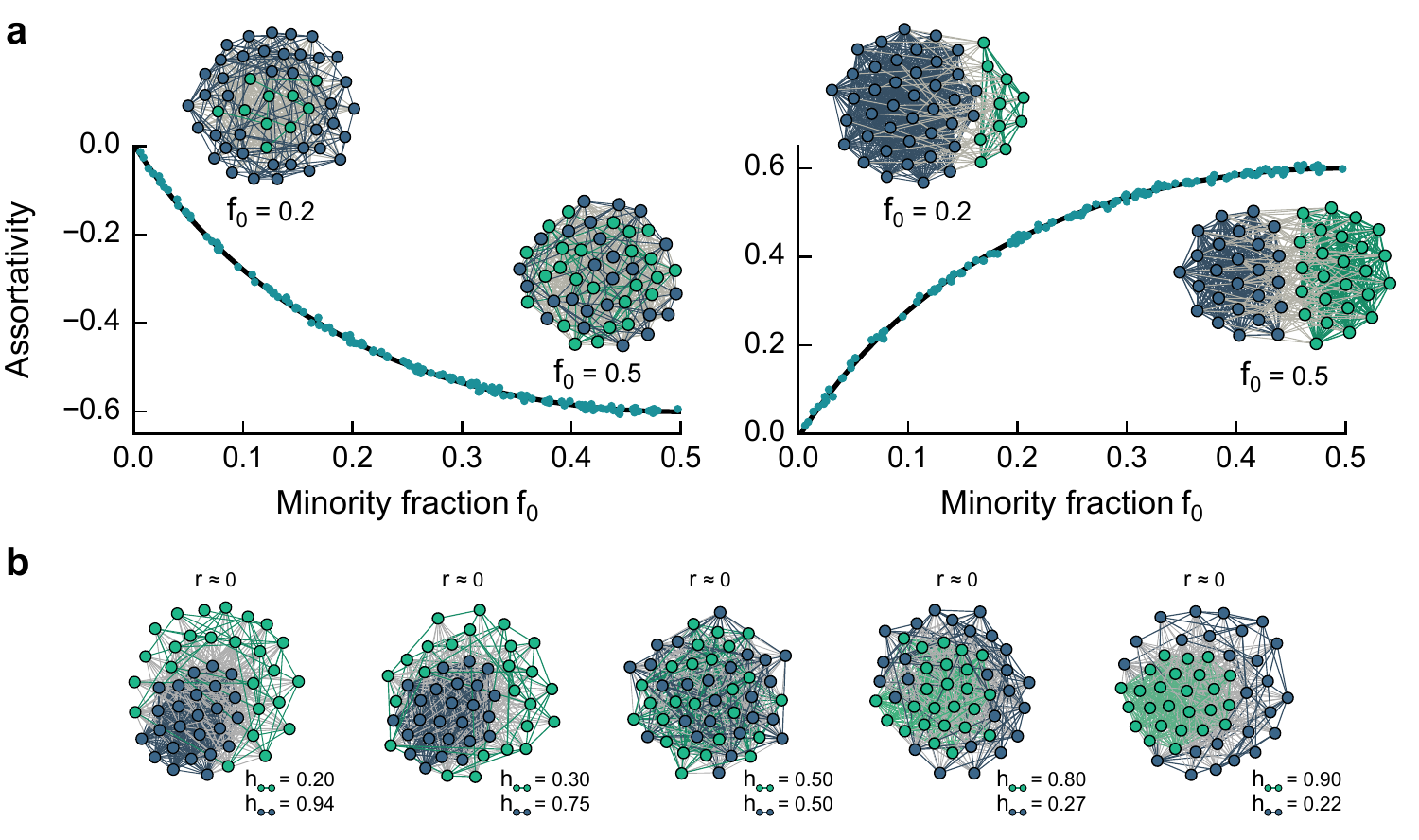}
\caption{\textbf{Nominal assortativity misses relevant mixing patterns in networks.}  \textbf{(a)} 
Nominal assortativity shows different mixing values for networks that have the same group mixing---a misrepresentation due to group-size imbalance. We generate these networks using a model with a group-mixing parameter $h$ that corresponds to the probability of same-group nodes being connected; the generated networks are in a heterophilic regime with $h=0.2$ (left) and a homophilic regime with $h=0.8$ (right). These networks have a fixed group mixing $h$ but varying minority fraction $f_0$. In the plots, solid lines represent the analytical formulation, whereas dots are values from simulations. 
%
\textbf{(b)} 
Nominal assortativity is a single-valued measure and ignores asymmetries in group mixing. The measure might indicate zero assortativity for networks with significant asymmetric mixing patterns. For example, a network with homophilic and heterophilic groups might be characterized with nominal assortativity equal to zero.}
\label{fig1}
\end{figure}

Here we demonstrate that nominal assortativity presents two fundamental inadequacies. First, it overlooks the group-size imbalance, implicitly assuming that groups are relatively equal. This assumption neglects that smaller groups have fewer possibilities to connect with themselves, misrepresenting mixing patterns in scenarios of group-size imbalance (see Fig.~\ref{fig1}A). Second, the measure consists of a uni-dimensional value, only characterizing symmetric group mixing (or an average mixing). This restriction ignores potential asymmetries in networks, thereby missing relevant mixing patterns (Fig.~\ref{fig1}B). Both inadequacies are problematic, particularly when analyzing real-world data and in the presence of minorities.

In real-world networks, groups tend to have unequal sizes, and some groups (i.e., minority groups) might be much smaller than the largest group. For instance, women are underrepresented in STEM fields, such as Computer Science and Physics, making them a minority group in professional networks\cite{national2019women,national2020,kong2022influence}. When analyzing such networks and other imbalanced data sets, we must consider group sizes to estimate the likelihood of in-group mixing biases. Besides unequal group sizes, networks might display asymmetries in how groups interact. For example, in male-dominated scientific fields, established researchers could be primarily men due to historical first-mover advantages \cite{kong2022influence}. Thus, senior men have the resources to drive their collaboration network, implying that the tendency for male-male collaboration may not be the same as female-female collaboration\cite{karimiminorities2022,jadidi2018gender}. In such settings, homophily is asymmetric, having different strengths for the minority and majority groups. These asymmetries, however, are lost when using a single-valued measure to characterize group mixing, such as nominal assortativity. 

In this work, we demonstrate how nominal assortativity misses relevant mixing patterns in networks with unequal group sizes or asymmetric mixing and we show how to tackle these shortcomings. First, we use generative network models with adjustable mixing parameters to show that nominal assortativity fails to recover the expected assortativity in synthetic networks. We characterize this limitation analytically and numerically by examining the relationship between assortativity, group size, and asymmetric mixing. Second, we propose the \textit{adjusted nominal assortativity} and show that this adjustment recovers the expected assortativity from synthetic networks. Third, we propose to assess asymmetric mixing in networks by estimating group-mixing tendencies using our analytical formulations. Finally, we discuss how our approach enables characterizing hidden mixing patterns in real-world networks.

\section*{Results}

Nominal assortativity (or discrete assortativity) characterizes mixing patterns by assessing the significance of the intra-group. To that end, this definition employs the $B \times B$ mixing matrix $\textbf{e}$ to account for groups connectivity, where $B$ is the number of groups, and each matrix element $e_{ij}$ corresponds to the fraction of edges connecting nodes from group $i$ to nodes from group $j$. The nominal assortativity measure is then defined as follows:
\begin{equation}
\label{eq:newman}
r = \dfrac{\sum_i e_{ii} - \sum_{i} a_i b_i}{1-\sum_i a_i b_i},
\end{equation}
where $a_{i}$ and $b_{i}$ are the fraction of edges that, respectively, begin and end at nodes from group $i$, defined as $a_{i} = \sum_{j} e_{ij}$ and $b_{i} = \sum_{j} e_{ji}$\cite{Newman2003mixing}. This definition produces an intuitive quantity that equals zero when groups lack intra- and inter-group tendencies (i.e., $e_{ii} = a_i b_i$). The quantity reaches to its maximum $r=1$ when intra-group ties dominate the network (i.e., $\sum_i e_{ii} = 1$) and becomes negative when inter-group ties are predominant. 


\subsection*{Nominal assortativity on networks with groups of unequal sizes }

To examine how nominal assortativity represents mixing patterns, we use generative network models in which we have a prior knowledge on what to expect from the value of mixing. Thus we aim to evaluate assortativity's ability to recover the expected mixing value. More precisely, we generate random networks using a model with a tunable group mixing parameter $h$, that corresponds to the probability of same-group nodes being connected, whereas its complement, $1-h$, is the probability of inter-group ties (see Methods). Here, we focus on the case of two groups, $B = 2$, in which nodes possess a binary attribute (e.g., red/blue, male/female). The case of beyond two groups is discussed in the supplementary materials. We examine networks with a fixed $h$ and varying group sizes, finding that nominal assortativity goes to zero as the minority group decreases in size (Fig.~\ref{fig1}A). For example, when $h=0.8$ (i.e., homophily), assortativity can vary from $0.6$ to $0$, depending on the proportional size of the minority group, despite fixed group mixing.

To investigate why nominal assortativity varies with the minority fraction, we turn to the analytical formulation of the assortativity. Let us use a more general notion of group mixing in which $h_{ii}$ denotes the intrinsic tendency of a node from group $i$ connecting to a node of the same group; its complement $h_{ij} = 1 - h_{ii}$ is the tendency of a node in group $i$ to connect to a node in group $j$. Therefore, in a random network, the probability of finding an edge between group $i$ and group $j$ express as $p_{ij} = f_if_jh_{ij}$, where $f$ corresponds to the proportional size of groups, implying that each mixing matrix element can be defined as $e_{ij} = p_{ij}/\sum_{ij} p_{ij}$, where the denominator is a normalizing factor. 

Thus, $\sum_i e_{ii}$ and $\sum_i a_i b_i$ can be expressed as follows:
\begin{equation}
\sum_i e_{ii} = \dfrac{{f_0}^2 h_{00}+ {f_1}^2 h_{11}}{\sum_{ij} p_{ij}},
\label{eq:sumeii_analytical}
\end{equation}
and
\begin{equation}
\sum_i a_i b_i = \dfrac{(f_0^2 h_{00} +  f_0 f_1 h_{01} )^2 + (f_1^2 h_{11} +  f_0 f_1  h_{10} )^2}{(\sum_{ij} p_{ij})^2},
\label{eq:sumaibi_analytical}
\end{equation}
where $0$ and $1$ are the labels for the minority and majority group, respectively. 
Finally, inserting Eq.~({\ref{eq:sumeii_analytical}}) and Eq.~({\ref{eq:sumaibi_analytical}}) into Eq.~({\ref{eq:newman}}), the nominal assortativity can be written as: 
\begin{equation}
\label{eq:r}
r = \dfrac{\dfrac{{f_0}^2 h_{00}+ {f_1}^2 h_{11}}{\sum_{ij} p_{ij}} - \dfrac{(f_0^2 h_{00} +  f_0 f_1 h_{01} )^2 + (f_1^2 h_{11} +  f_0 f_1  h_{10} )^2}{(\sum_{ij} p_{ij})^2}}{1-\dfrac{( f_0^2 h_{00} +  f_0 f_1 h_{01} )^2 + ( f_1^2 h_{11} +  f_0 f_1  h_{10})^2}{(\sum_{ij} p_{ij})^2}}.
\end{equation}

\begin{figure}[b!]
\centering
\begin{adjustwidth}{-.4in}{0in} 
\includegraphics[width=7in]{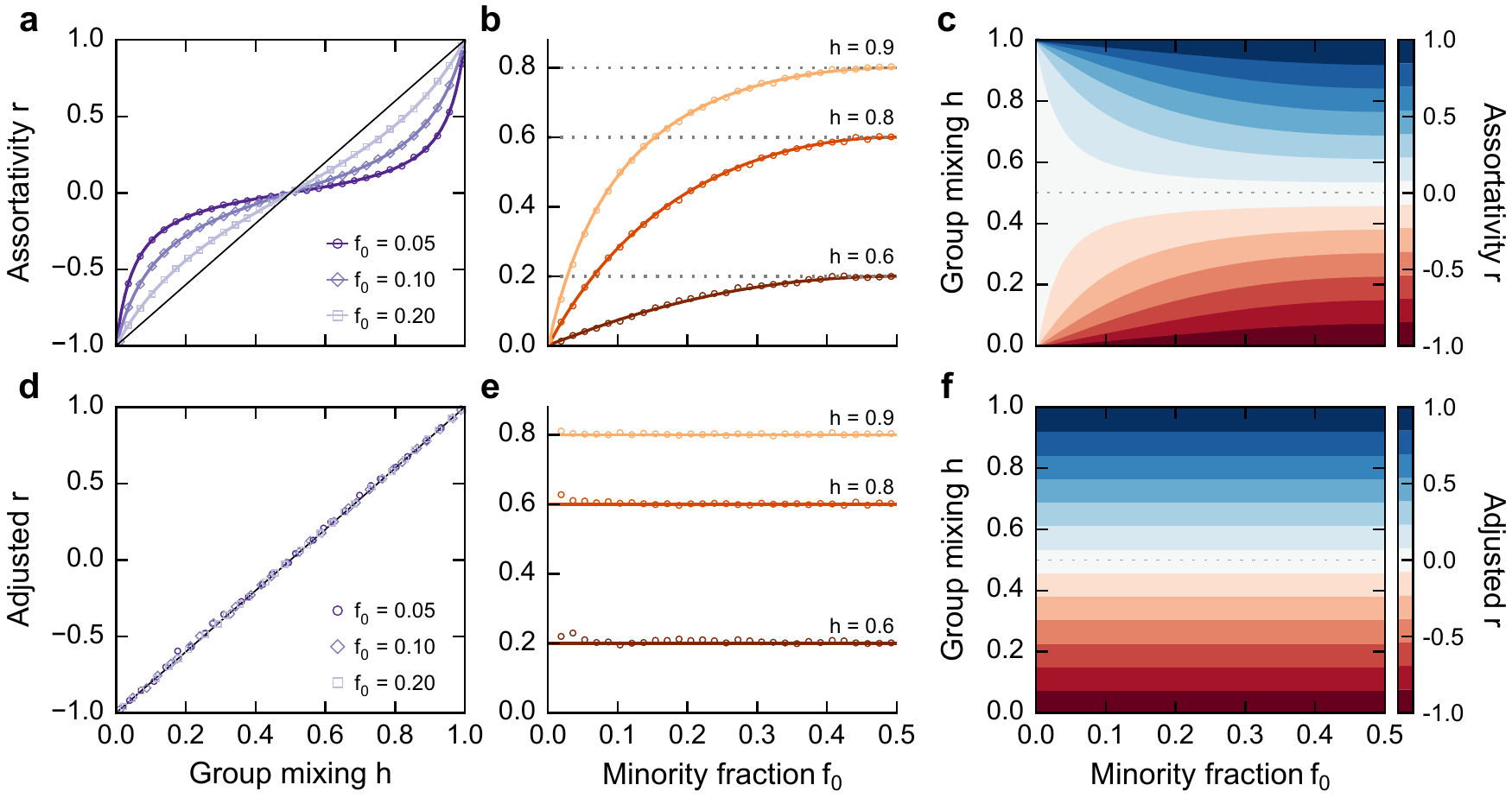}
\end{adjustwidth}
\caption{\textbf{Adjusted assortativity retrieves the expected assortativity in networks with group-size imbalance.} 
\textbf{(a)} Nominal assortativity has a group-size dependence that \textbf{(b)} underestimates the strength of group mixing in networks. 
\textbf{(c)} This underestimation is more severe in the presence of smaller minority groups. \textbf{(d)} We propose the \textit{adjusted assortativity} that tackles this misrepresentation by adjusting for group sizes in the network. \textbf{(e)} The measure has a linear relationship with group mixing $h$ and \textbf{(f)} is independent of group sizes. In all plots, solid lines represent the analytical formulation, whereas dots are values from simulations. 
%
}
\label{fig:h_adj_assort_ER_sym}
\end{figure}

This equation reveals that nominal assortativity is a function of group sizes $f_0$ and $f_1$. We verify this group-size dependency by comparing our analytical formulation with the assortativity measured on synthetic networks, finding a perfect agreement between Eq.~(\ref{eq:r}) and simulations (Fig.~\ref{fig:h_adj_assort_ER_sym}A-B). Our results confirm the group-size dependence and reveal that this dependence increases with smaller minority groups (Fig.~\ref{fig:h_adj_assort_ER_sym}C). In contrast, when groups have similar sizes, we observe, as expected, a linear relationship between group mixing $h$ and nominal assortativity. More precisely, when groups are equal in size, $f_0 = f_1 = 0.5$, Eq.~(\ref{eq:r}) becomes $r = h_{00} + h_{11} - 1$. The group-size dependency occurs in other types of networks such as scale-free networks. For instance, we simulate the Barab\'{a}si--Albert homophily (BA-Homophily) model, which incorporates group mixing preferences with the preferential attachment \cite{karimi2018homophily}, and demonstrate that nominal assortativity is a function of group sizes on such networks and in scenarios involving more than two groups (see Supplementary Material). Overall, these findings imply that nominal assortativity is unadjusted for group sizes and introduces an artifactual bias into mixing analyses in imbalanced scenarios. 

\subsection*{The adjusted nominal assortativity}
Here we propose to adjust the nominal assortativity for group sizes by normalizing the elements of the mixing matrix. This approach accurately retrieves the expected assortativity in networks, enabling us to assess mixing patterns in imbalanced networks. To that end, we define the \textit{adjusted mixing matrix} $\mathbf{e^\star}$, which accounts for the network's pool of opportunities, namely, the fact that larger groups have more opportunities to connect and thus should be normalized by the group size. We define each element of the adjusted mixing matrix $\mathbf{e^\star}$ to be
\begin{linenomath}
    \begin{equation*}
        e^\star_{ij} = \dfrac{e_{ij}}{f_if_j},
    \end{equation*}
\end{linenomath}
where $f_k$ corresponds to the proportional size of group $k$. This adjustment, ensures that the elements of the mixing matrix only represents the mixing tendencies ($h$) that are relevant for measuring intrinsic homophily and assortativity and not other factors. For instance, in the case of two groups, where original mixing elements are $e_{00} \simeq f_0^2 h_{00}$ and  $e_{11} \simeq f_1^2 h_{11}$, the adjusted elements of the matrix are expressed as $e_{00}^\star \simeq h_{00}$ and  $e_{11}^\star \simeq h_{11}$. 

Moreover, we define the \textit{adjusted nominal assortativity}, $r_{adj}$, as follows: 
\begin{linenomath}
    \begin{equation*}
        r_{adj} = \dfrac{\sum_i e^\star_{ii} - \sum_{i} a^\star_i b^\star_i}{1-\sum_i a^\star_i b^\star_i},
    \end{equation*}
\end{linenomath}
where $a^\star_{i} = \sum_{j} e^\star_{ij}$ and $b_{i} = \sum_{j} e^\star_{ji}$. This adjustment considers the effects of group-size imbalance on the mixing matrix, leading to a consistent assessment of mixing patterns in imbalanced scenarios. 

We verify the proposed measure by generating synthetic data with different group-imbalance and mixing scenarios. We examine networks generated with a fixed $h$ and varying group sizes, revealing that adjusted nominal assortativity accurately recovers the expected mixing independent of group sizes (Fig.~\ref{fig:h_adj_assort_ER_sym}D-E). Thus results show that the adjusted nominal assortativity has a linear relationship with group mixing $h$, regardless of $f_0$ and $f_1$ as expected (see Fig.~\ref{fig:h_adj_assort_ER_sym}F). We find similar results for scale-free networks and three-groups scenarios (see Supplementary Material). In sum, the adjusted nominal assortativity accounts for group sizes and pool of opportunities, enabling us to assess group mixing preferences accurately. 

\subsection*{Assessing group mixing in empirical networks}

Next, we explore nominal assorativity in different real-world networks with unequal group sizes. Results show that nominal assortativity underestimates the mixing patterns compared to the adjusted nominal assortativity (see Table~\ref{tabledata}). We analyze social networks of academic collaboration and face-to-face interactions with annotated binary gender information (see Supplementary Material for detailed data descriptions)~\cite{Fournet2014,Mastrandrea2015,genois2022combining}. In most cases, assortativity $r$ is lower than the adjusted assortativity $r_{adj}$, especially in the cases of small minority groups. For example, in the collaborative coding platform GitHub, where women are only 6\% of the network, nominal assortativity is $r=0.04$, implying the absence of assortative collaboration; in contrast, the adjusted assortativity is $r_{adj}=0.16$, suggesting a potential gender assortativity. Similarly, nominal assortativity might mislead us to mistake changes in group mixing for changes in group sizes. For instance, in the collaboration network among computer scientists (DBLP), assortativity $r$ increases from $r=0.04$ in 1980 to $r=0.10$ in 2010, which could imply a possible change in group mixing over time. However, this change might be merely due to the growth of the minority group. The minority size almost doubled, from 11\% to 21\%, and the adjusted assortativity indicates a stable mixing around $r_{adj} = 0.15$. Overall, these findings underscore the importance of accounting for group sizes when analyzing mixing patterns and the risks of ignoring group imbalance in networks.

\begin{table}[h!]
\caption{\textbf{Nominal assortativity and adjusted assortativity in empirical networks.} $N$ denotes number of nodes, $f_0$ is the minority fraction, $E$ is the total number of edges, and label $0$ refers to the minority group and label $1$ refers to the majority group.\label{tabledata}}
\centering
\begin{tabular}{lrrrrrrrr}
\toprule
Network     & \multicolumn{1}{c}{$N$} & \multicolumn{1}{c}{$f_0$} & \multicolumn{1}{c}{$E$} &
\multicolumn{1}{c}{$E_{00}$} &
\multicolumn{1}{c}{$E_{11}$} &
\multicolumn{1}{c}{$E_{01}$} &
\multicolumn{1}{c}{$r$} & \multicolumn{1}{c}{$r_{adj}$}     \\\midrule
APS (2000) &       8,285 & 0.11 &       9,071 &        126 &       1,539 &       7,406 & 0.05 & 0.11 \\
GitHub  &     311,755 & 0.06 &    1,537,570 &       7,432 &     149,069 &    1,381,069 & 0.04 & 0.15 \\
DBLP (2010) &     170,984 & 0.21 &     322,052 &      17,468 &      91,738 &     212,846 & 0.10 & 0.14 \\
DBLP (2000) &      54,966 & 0.18 &      72,369 &       3,123 &      18,101 &      51,145 & 0.11 & 0.16 \\
DBLP (1990) &      13,764 & 0.15 &      12,178 &        384 &       2,701 &       9,093 & 0.09 & 0.16 \\
DBLP (1980) &       2,664 & 0.11 &       1,765 &         24 &        274 &       1,467 & 0.06 & 0.16 \\
INFORMS (2010) &       1,426 & 0.16 &       1,009 &         34 &        247 &        728 & 0.07 & 0.12 \\
SocioPatterns 4 &        180 & 0.27 &       2,220 &        182 &        762 &       1,276 & 0.09 & 0.12 \\
SocioPatterns 5 &        327 & 0.44 &       5,818 &       1,471 &       2348 &       1,999 & 0.19 & 0.19 \\
\bottomrule 
\end{tabular}
\end{table}

\subsection*{Mixing patterns in networks with asymmetric mixing}


A single measure of assortativity reduces information about the $B\times B$ mixing matrix into a single value, leading to a concise measure but potentially missing relevant asymmetries in mixing. The idea that a single summary statistic may not be representative of a dataset is, of course, not new and has been shown in prior works \cite{anscombe1973graphs}. More recently, Peel and colleagues showed the heterogeneity of the local assortativity \cite{peel2018multiscale} and Piraveenan et al.~\cite{piraveenan2012congruity} showed the extent to which each node contributes to the measure of assortativity.  Here, we pay a special attention to the asymmetric nature of group mixing while assuming no heterogeneity at the node level.

\begin{figure}[b!]
\centering
\begin{adjustwidth}{-.4in}{0in} 
\includegraphics[width=6.92in]{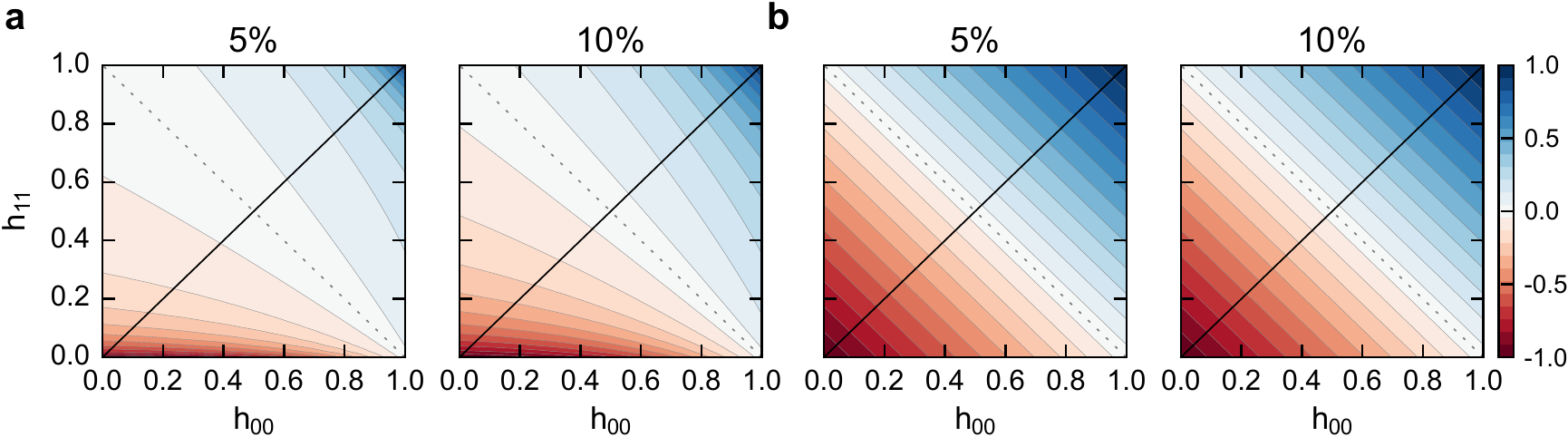}
\end{adjustwidth}
\caption{\textbf{Unidimensional measures of assortativity overlook asymmetry in networks. } (\textbf{a}) The nominal assortativity is dependent of group size in asymmetric cases, whereas (\textbf{b}) the adjusted version is size-independent. Yet both versions of assortativity ignore asymmetric mixing; they reduce a mixing matrix into a unidimensional value, producing a concise measure but missing asymmetry in networks. These measures might indicate an absence of mixing tendency despite significant group mixing. In particular, both measures are zero when $h_{00} = 1 - h_{11}$ (i.e., the dashed lines). In the plots, each heatmap displays the respective measures in varying mixing scenarios of minority mixing $h_{00}$ and majority mixing $h_{11}$ in the presence of minority sizes $f_0=0.05$ and $f_0=0.10$.
%
%
}  
\label{fig:h_adj_assort_asym_ER}
\end{figure}

To characterize $r$ and $r_{adj}$ in asymmetric scenarios, we relax the assumption of $h_{00}=h_{11}=h$ and use our analytical formulation (Eq.~(\ref{eq:r})) to evaluate the nominal assortativity over the whole parameter space of $h_{00}$ and $h_{11}$ (see Fig.~\ref{fig:h_adj_assort_asym_ER}). We find that the adjusted nominal assortativity is consistent and independent of group size in asymmetric cases, whereas the unadjusted version is size-dependent. Both measures, however, might characterize contrasting mixing patterns with the same value. In particular, these measures might indicate an absence of inter- or intra- group mixing tendency despite significant group mixing. For instance, when $h_{00} = 0.8$ and $h_{11} = 0.2$, the minority group has a strong homophilic tendency, whereas the majority has a strong heterophilic tendency; yet, nominal assortativity equals zero, incorrectly suggesting a lack of assortative or disassortative patterns (Fig.~\ref{fig:h_adj_assort_asym_ER}). 

To better understand the underlying reason for this misrepresentation, note that when $r=0$, the numerator in Eq.~(\ref{eq:r}) is zero, leading to the following equation:
\begin{linenomath}
    \begin{equation*}
    \label{eq:zeror}
    \dfrac{{f_0}^2 h_{00}+ {f_1}^2 h_{11}}{\sum_{ij} p_{ij}} - \dfrac{(f_0^2 h_{00} +  f_0 f_1 h_{01} )^2 + (f_1^2 h_{11} +  f_0 f_1  h_{10} )^2}{(\sum_{ij} p_{ij})^2} = 0.
    \end{equation*}
\end{linenomath}
Simplifying this equation by using the expression of Eq.~\ref{eq:sum_p}, we find that $h_{00} = 1 - h_{11}$ satisfies this condition. In other words, in many cases when nominal assortavitiy reports a zero value (i.e., lack of any dis/assortative preferences), the group mixing tendencies could be widely different. These findings show that compressing the mixing matrix into a single value, such as assortativity, can hide relevant asymmetric mixing patterns that are present in networks. It is worth noting that paying attention to asymmetries in mixing patterns between groups is important in other applications, such as the emergence of core-periphery structures \cite{urena2023assortative}. 

\subsection*{Assessing asymmetric mixing patterns in networks}

In order to assess asymmetric mixing among groups, we propose to turn to  the mixing probabilities in a network given an assumption of its generative process. For example, in a random homophilic networks described earlier (ER-Homophily), the diagonal of the mixing matrix can be expressed as:

\begin{linenomath}
    \begin{equation*}
    e_{00} = \frac{{f_0}^2 h_{00}}{\sum_{ij} p_{ij}} \quad\text{and}\quad e_{11} = \frac{{f_1}^2 h_{11}}{\sum_{ij} p_{ij}},
    \end{equation*}
\end{linenomath}
which can be re-written as follows:
\begin{equation}
\label{eq:h_ER}
h_{00} = \frac{E_{00}}{E}\frac{\sum_{ij} p_{ij}}{{f_0}^2} \quad\text{and}\quad  h_{11} = \frac{E_{11}}{E}\frac{\sum_{ij} p_{ij}}{{f_1}^2},
\end{equation}
where
\begin{linenomath}
    \begin{equation}
    \label{eq:sum_p}
    \sum_{ij} p_{ij} = {f_0}^2 h_{00} +  {f_1}^2 h_{11} + {f_0}{f_1}h_{01} + {f_0}{f_1}h_{10},
    \end{equation}
\end{linenomath}
and $E$ is the total number of edges, and $E_{00}$ and $E_{11}$ are the number of intra-group edges of the minority and majority groups, and $e_{00}$ and $e_{11}$ are fraction of intra-group edges normalized by $E$.
By combining the equations above, $\sum_{ij} p_{ij}$ can be expressed as:
\begin{linenomath}
\begin{equation}
    \label{eq:sum_ER}
    \sum_{ij} p_{ij} = \frac{2 f_0 f_1}{1-e_{00}(1-f_1/f_0) - e_{11}(1-f_0/f_1)}.
\end{equation}
\end{linenomath}
 By using Eq.~(\ref{eq:h_ER}) and Eq.~(\ref{eq:sum_ER}), we can retrieve $h_{00}$ and $h_{11}$ from data, given we know basic information about the network (i.e., $E$, $E_{00}$, $E_{11}$, and $f_0$). 

We verify this method by generating networks with varying mixing parameters and compare the estimated parameters with the ground truth in Supplementary Note 6. Similar methodology can be applied to scale-free networks, finding equivalent results (see Supplementary Note 6). Though this approach requires prior knowledge about the underlying generative processes in networks, it is plausible to argue that many small-scale and large-scale social networks often fall into these two categories of topologically random or scale-free structure \cite{Isella2011,broido2019scale,holme2019rare}. Once the plausible topology is identified by examining the degree distribution, the appropriate formulation can be used to extract the group mixing asymmetries.




\section*{Discussion}
Despite its popularity and relative accuracy in capturing homophily and assortative mixing in a variety of networks, nominal assortativity can produce distorted assessments of mixing patterns in networks with unequal groups and asymmetric mixing. In this work, we demonstrated this problem and proposed ways to tackle these limitations. Using generative network models with adjustable mixing, we show that nominal assortativity fails to estimate assortativity in networks accurately. Our results demonstrate (1)~the need for accounting for group sizes in mixing analyses and (2)~the inability of single-valued measures to capture asymmetries in networks. 

To tackle these limitations, we develop two approaches to assess group mixing in networks. First, we propose adjusted nominal assortativity to solve the group-size limitation, which accurately recovers the expected assortativity from networks. Our analysis of real-world networks reveals that nominal assortativity underestimates the strength of mixing patterns compared to the adjusted assortativity. Second, we propose to assess asymmetric mixing in networks by estimating the intra-group mixing probabilities accounting for group size differences and other group-level topological features. It is worth mentioning that there are a variety of other segregation and assortativity measurements in the social network literature beyond the nominal assortativity. Future works should focus on comparing the sensitivity, equivalency, and compatibility of those measurements against each other and baseline scenarios similar to this paper and the previous efforts \cite{bojanowski2014measuring}. 

Accurately measuring biases in group mixing in social networks is crucial because mixing biases affect perception of minorities \cite{Lee2019}, access to social capital \cite{jackson2021inequality}, and algorithmic visibility \cite{espin2022inequality}, to name a few. Our work lays a novel foundation by proposing an accurate measure of assortativity that can be applied to a wide range of social networks. Better assessment of group-level tendencies and asymmetries in networks provides the means to understand how diverse groups interact---a fundamental step for uncovering mechanisms governing our social lives.  

\section*{Methods}

\subsection*{Random networks with group mixing}
To analyze assortativity in networks, we use a model that incorporates group mixing and random tie formation in networks. In this model, an edge between two nodes depends on their group memberships via a stochastic process. The probability of a node from group $i$ to establish a tie with a node from group $j$ is denoted as $h_{ij}$. The probability to connect with nodes of the same group is thus the complementary function so that $h_{ij} = 1 - h_{ii}$, likewise $h_{ji} = 1 - h_{jj}$ (see Jupyter notebooks for more details).





\bibliographystyle{naturemag}
\bibliography{references}

\begin{thebibliography}{10}
\expandafter\ifx\csname url\endcsname\relax
  \def\url#1{\texttt{#1}}\fi
\expandafter\ifx\csname urlprefix\endcsname\relax\def\urlprefix{URL }\fi
\providecommand{\bibinfo}[2]{#2}
\providecommand{\eprint}[2][]{\url{#2}}

\bibitem{McPherson2001}
\bibinfo{author}{McPherson, M.}, \bibinfo{author}{Smith-Lovin, L.} \&
  \bibinfo{author}{Cook, J.~M.}
\newblock \bibinfo{title}{{Birds of a Feather: Homophily in Social Networks}}.
\newblock \emph{\bibinfo{journal}{Annual Review of Sociology}}
  \textbf{\bibinfo{volume}{27}}, \bibinfo{pages}{415--444}
  (\bibinfo{year}{2001}).

\bibitem{goodreau2009birds}
\bibinfo{author}{Goodreau, S.~M.}, \bibinfo{author}{Kitts, J.~A.} \&
  \bibinfo{author}{Morris, M.}
\newblock \bibinfo{title}{Birds of a feather, or friend of a friend? using
  exponential random graph models to investigate adolescent social networks}.
\newblock \emph{\bibinfo{journal}{Demography}} \textbf{\bibinfo{volume}{46}},
  \bibinfo{pages}{103--125} (\bibinfo{year}{2009}).

\bibitem{oliveira2022group}
\bibinfo{author}{Oliveira, M.} \emph{et~al.}
\newblock \bibinfo{title}{Group mixing drives inequality in face-to-face
  gatherings}.
\newblock \emph{\bibinfo{journal}{Communications Physics}}
  \textbf{\bibinfo{volume}{5}}, \bibinfo{pages}{1--9} (\bibinfo{year}{2022}).

\bibitem{Newman2003mixing}
\bibinfo{author}{Newman, M. E.~J.}
\newblock \bibinfo{title}{{Mixing patterns in networks}}.
\newblock \emph{\bibinfo{journal}{Physical Review E}}
  \textbf{\bibinfo{volume}{67}}, \bibinfo{pages}{026126}
  (\bibinfo{year}{2003}).

\bibitem{peel2018multiscale}
\bibinfo{author}{Peel, L.}, \bibinfo{author}{Delvenne, J.-C.} \&
  \bibinfo{author}{Lambiotte, R.}
\newblock \bibinfo{title}{Multiscale mixing patterns in networks}.
\newblock \emph{\bibinfo{journal}{Proceedings of the National Academy of
  Sciences}} \textbf{\bibinfo{volume}{115}}, \bibinfo{pages}{4057--4062}
  (\bibinfo{year}{2018}).

\bibitem{schaible2021sensing}
\bibinfo{author}{Schaible, J.}, \bibinfo{author}{Oliveira, M.},
  \bibinfo{author}{Zens, M.} \& \bibinfo{author}{G{\'e}nois, M.}
\newblock \bibinfo{title}{Sensing close-range proximity for studying
  face-to-face interaction}.
\newblock In \emph{\bibinfo{booktitle}{Handbook of Computational Social
  Science, Volume 1}}, \bibinfo{pages}{219--239}
  (\bibinfo{publisher}{Routledge}, \bibinfo{year}{2021}).

\bibitem{efferson2008coevolution}
\bibinfo{author}{Efferson, C.}, \bibinfo{author}{Lalive, R.} \&
  \bibinfo{author}{Fehr, E.}
\newblock \bibinfo{title}{The coevolution of cultural groups and ingroup
  favoritism}.
\newblock \emph{\bibinfo{journal}{Science}} \textbf{\bibinfo{volume}{321}},
  \bibinfo{pages}{1844--1849} (\bibinfo{year}{2008}).

\bibitem{Traud2012}
\bibinfo{author}{Traud, A.~L.}, \bibinfo{author}{Mucha, P.~J.} \&
  \bibinfo{author}{Porter, M.~A.}
\newblock \bibinfo{title}{{Social structure of Facebook networks}}.
\newblock \emph{\bibinfo{journal}{Physica A: Statistical Mechanics and its
  Applications}} \textbf{\bibinfo{volume}{391}}, \bibinfo{pages}{4165--4180}
  (\bibinfo{year}{2012}).

\bibitem{Pepe2010}
\bibinfo{author}{Pepe, A.} \& \bibinfo{author}{Rodriguez, M.~A.}
\newblock \bibinfo{title}{{Collaboration in sensor network research: an
  in-depth longitudinal analysis of assortative mixing patterns}}.
\newblock \emph{\bibinfo{journal}{Scientometrics}}
  \textbf{\bibinfo{volume}{84}}, \bibinfo{pages}{687--701}
  (\bibinfo{year}{2010}).

\bibitem{Jacobson2009}
\bibinfo{author}{Jacobson, K.} \& \bibinfo{author}{Sandler, M.}
\newblock \bibinfo{title}{{Musically Meaningful or Just Noise? An Analysis of
  On-line Artist Networks}}.
\newblock In \emph{\bibinfo{booktitle}{Lecture Notes in Computer Science
  (including subseries Lecture Notes in Artificial Intelligence and Lecture
  Notes in Bioinformatics)}}, vol. \bibinfo{volume}{5493 LNCS},
  \bibinfo{pages}{107--118} (\bibinfo{year}{2009}).

\bibitem{Oliveira2014}
\bibinfo{author}{Oliveira, M.}, \bibinfo{author}{Bastos-Filho, C.} \&
  \bibinfo{author}{Menezes, R.}
\newblock \bibinfo{title}{{Political Social Networks Reveal Strong Party
  Loyalty in Brazil and Weak Regionalism}}.
\newblock In \emph{\bibinfo{booktitle}{The Sixth ASE International Conference
  on Social Computing}}, \bibinfo{pages}{1--8} (\bibinfo{address}{Stanford,
  USA}, \bibinfo{year}{2014}).
\newblock \urlprefix\url{http://doi.org/10.13140/RG.2.2.35595.08489}.

\bibitem{Bucur2019}
\bibinfo{author}{Bucur, D.}
\newblock \bibinfo{title}{{Gender homophily in online book networks}}.
\newblock \emph{\bibinfo{journal}{Information Sciences}}
  \textbf{\bibinfo{volume}{481}}, \bibinfo{pages}{229--243}
  (\bibinfo{year}{2019}).

\bibitem{cells_network}
\bibinfo{author}{Cheng, S.}, \bibinfo{author}{Price, D.~C.},
  \bibinfo{author}{Karkar, S.} \& \bibinfo{author}{Bhattacharya, D.}
\newblock \bibinfo{title}{Exploring biotic interactions within protist cell
  populations using network methods}.
\newblock \emph{\bibinfo{journal}{Journal of Eukaryotic Microbiology}}
  \textbf{\bibinfo{volume}{61}}, \bibinfo{pages}{399--403}
  (\bibinfo{year}{2014}).

\bibitem{Lusseau2004}
\bibinfo{author}{Lusseau, D.} \& \bibinfo{author}{Newman, M. E.~J.}
\newblock \bibinfo{title}{{Identifying the role that animals play in their
  social networks}}.
\newblock \emph{\bibinfo{journal}{Proceedings of the Royal Society of London.
  Series B: Biological Sciences}} \textbf{\bibinfo{volume}{271}},
  \bibinfo{pages}{477--481} (\bibinfo{year}{2004}).

\bibitem{cinelli2020network}
\bibinfo{author}{Cinelli, M.}, \bibinfo{author}{Peel, L.},
  \bibinfo{author}{Iovanella, A.} \& \bibinfo{author}{Delvenne, J.-C.}
\newblock \bibinfo{title}{Network constraints on the mixing patterns of binary
  node metadata}.
\newblock \emph{\bibinfo{journal}{Physical Review E}}
  \textbf{\bibinfo{volume}{102}}, \bibinfo{pages}{062310}
  (\bibinfo{year}{2020}).

\bibitem{national2019women}
\bibinfo{author}{{National Science Foundation, National Center for Science and
  Engineering Statistics}}.
\newblock \emph{\bibinfo{title}{Women, Minorities, and Persons with
  Disabilities in Science and Engineering: 2019. Special Report NSF 19-304.}}
  (\bibinfo{publisher}{ERIC Clearinghouse}, \bibinfo{year}{2019}).

\bibitem{national2020}
\bibinfo{author}{{National Science Board, National Science Foundation}}.
\newblock \emph{\bibinfo{title}{Science and Engineering Indicators 2020: The
  State of U.S. Science and Engineering. NSB-2020-1.}}
  (\bibinfo{publisher}{ERIC Clearinghouse}, \bibinfo{year}{2020}).

\bibitem{kong2022influence}
\bibinfo{author}{Kong, H.}, \bibinfo{author}{Martin-Gutierrez, S.} \&
  \bibinfo{author}{Karimi, F.}
\newblock \bibinfo{title}{Influence of the first-mover advantage on the gender
  disparities in physics citations}.
\newblock \emph{\bibinfo{journal}{Communications Physics}}
  \textbf{\bibinfo{volume}{5}}, \bibinfo{pages}{1--11} (\bibinfo{year}{2022}).

\bibitem{karimiminorities2022}
\bibinfo{author}{Karimi, F.}, \bibinfo{author}{Oliveira, M.} \&
  \bibinfo{author}{Strohmaier, M.}
\newblock \bibinfo{title}{Minorities in networks and algorithms}.
\newblock \emph{\bibinfo{journal}{arXiv preprint arXiv:2206.07113}}
  (\bibinfo{year}{2022}).

\bibitem{jadidi2018gender}
\bibinfo{author}{Jadidi, M.}, \bibinfo{author}{Karimi, F.},
  \bibinfo{author}{Lietz, H.} \& \bibinfo{author}{Wagner, C.}
\newblock \bibinfo{title}{Gender disparities in science? dropout, productivity,
  collaborations and success of male and female computer scientists}.
\newblock \emph{\bibinfo{journal}{Advances in Complex Systems}}
  \textbf{\bibinfo{volume}{21}}, \bibinfo{pages}{1750011}
  (\bibinfo{year}{2018}).

\bibitem{karimi2018homophily}
\bibinfo{author}{Karimi, F.}, \bibinfo{author}{G{\'e}nois, M.},
  \bibinfo{author}{Wagner, C.}, \bibinfo{author}{Singer, P.} \&
  \bibinfo{author}{Strohmaier, M.}
\newblock \bibinfo{title}{Homophily influences ranking of minorities in social
  networks}.
\newblock \emph{\bibinfo{journal}{Scientific reports}}
  \textbf{\bibinfo{volume}{8}}, \bibinfo{pages}{11077} (\bibinfo{year}{2018}).

\bibitem{Fournet2014}
\bibinfo{author}{Fournet, J.} \& \bibinfo{author}{Barrat, A.}
\newblock \bibinfo{title}{{Contact Patterns among High School Students}}.
\newblock \emph{\bibinfo{journal}{PLoS ONE}} \textbf{\bibinfo{volume}{9}},
  \bibinfo{pages}{e107878} (\bibinfo{year}{2014}).

\bibitem{Mastrandrea2015}
\bibinfo{author}{Mastrandrea, R.}, \bibinfo{author}{Fournet, J.} \&
  \bibinfo{author}{Barrat, A.}
\newblock \bibinfo{title}{Contact patterns in a high school: A comparison
  between data collected using wearable sensors, contact diaries and friendship
  surveys}.
\newblock \emph{\bibinfo{journal}{PLoS ONE}} \textbf{\bibinfo{volume}{10}},
  \bibinfo{pages}{e0136497} (\bibinfo{year}{2015}).

\bibitem{genois2022combining}
\bibinfo{author}{G{\'e}nois, M.} \emph{et~al.}
\newblock \bibinfo{title}{Combining sensors and surveys to study social
  contexts: Case of scientific conferences}.
\newblock \emph{\bibinfo{journal}{arXiv preprint arXiv:2206.05201}}
  (\bibinfo{year}{2022}).

\bibitem{anscombe1973graphs}
\bibinfo{author}{Anscombe, F.~J.}
\newblock \bibinfo{title}{Graphs in statistical analysis}.
\newblock \emph{\bibinfo{journal}{The american statistician}}
  \textbf{\bibinfo{volume}{27}}, \bibinfo{pages}{17--21}
  (\bibinfo{year}{1973}).

\bibitem{piraveenan2012congruity}
\bibinfo{author}{Piraveenan, M.}, \bibinfo{author}{Prokopenko, M.} \&
  \bibinfo{author}{Zomaya, A.~Y.}
\newblock \bibinfo{title}{On congruity of nodes and assortative information
  content in complex networks.}
\newblock \emph{\bibinfo{journal}{Networks Heterog. Media}}
  \textbf{\bibinfo{volume}{7}}, \bibinfo{pages}{441--461}
  (\bibinfo{year}{2012}).

\bibitem{urena2023assortative}
\bibinfo{author}{Ure{\~n}a-Carrion, J.}, \bibinfo{author}{Karimi, F.},
  \bibinfo{author}{I{\~n}iguez, G.} \& \bibinfo{author}{Kivel{\"a}, M.}
\newblock \bibinfo{title}{Assortative and preferential attachment lead to
  core-periphery networks}.
\newblock \emph{\bibinfo{journal}{arXiv preprint arXiv:2305.15061}}
  (\bibinfo{year}{2023}).

\bibitem{Isella2011}
\bibinfo{author}{Isella, L.} \emph{et~al.}
\newblock \bibinfo{title}{{Close Encounters in a Pediatric Ward: Measuring
  Face-to-Face Proximity and Mixing Patterns with Wearable Sensors}}.
\newblock \emph{\bibinfo{journal}{PLoS ONE}} \textbf{\bibinfo{volume}{6}},
  \bibinfo{pages}{e17144} (\bibinfo{year}{2011}).

\bibitem{broido2019scale}
\bibinfo{author}{Broido, A.~D.} \& \bibinfo{author}{Clauset, A.}
\newblock \bibinfo{title}{Scale-free networks are rare}.
\newblock \emph{\bibinfo{journal}{Nature communications}}
  \textbf{\bibinfo{volume}{10}}, \bibinfo{pages}{1017} (\bibinfo{year}{2019}).

\bibitem{holme2019rare}
\bibinfo{author}{Holme, P.}
\newblock \bibinfo{title}{Rare and everywhere: Perspectives on scale-free
  networks}.
\newblock \emph{\bibinfo{journal}{Nature communications}}
  \textbf{\bibinfo{volume}{10}}, \bibinfo{pages}{1016} (\bibinfo{year}{2019}).

\bibitem{bojanowski2014measuring}
\bibinfo{author}{Bojanowski, M.} \& \bibinfo{author}{Corten, R.}
\newblock \bibinfo{title}{Measuring segregation in social networks}.
\newblock \emph{\bibinfo{journal}{Social networks}}
  \textbf{\bibinfo{volume}{39}}, \bibinfo{pages}{14--32}
  (\bibinfo{year}{2014}).

\bibitem{Lee2019}
\bibinfo{author}{Lee, E.} \emph{et~al.}
\newblock \bibinfo{title}{{Homophily and minority-group size explain perception
  biases in social networks}}.
\newblock \emph{\bibinfo{journal}{Nature Human Behaviour}}
  \textbf{\bibinfo{volume}{3}}, \bibinfo{pages}{1078--1087}
  (\bibinfo{year}{2019}).

\bibitem{jackson2021inequality}
\bibinfo{author}{Jackson, M.~O.}
\newblock \bibinfo{title}{Inequality's economic and social roots: The role of
  social networks and homophily}.
\newblock \emph{\bibinfo{journal}{Available at SSRN 3795626}}
  (\bibinfo{year}{2021}).

\bibitem{espin2022inequality}
\bibinfo{author}{Esp{\'\i}n-Noboa, L.}, \bibinfo{author}{Wagner, C.},
  \bibinfo{author}{Strohmaier, M.} \& \bibinfo{author}{Karimi, F.}
\newblock \bibinfo{title}{Inequality and inequity in network-based ranking and
  recommendation algorithms}.
\newblock \emph{\bibinfo{journal}{Scientific reports}}
  \textbf{\bibinfo{volume}{12}}, \bibinfo{pages}{2012} (\bibinfo{year}{2022}).

\end{thebibliography}


\section*{Data Availability}
The sources of all empirical data used in our analyses are described in Supplementary Note~1. 

\section*{Code Availability}
All relevant code used in this study will be available at \url{https://github.com/macoj/assortativity}. 

\section*{Ethics declarations}
\subsection*{Competing interests}
 The authors declare no competing interests.
 
 \section*{Authors Contributions}
F.K. and M.O. proposed the project, wrote relevant code, carried out analytical analyses, wrote the first draft, and reviewed the final manuscript.

\section*{Acknowledgments} 
We thank Leto Peel, Matteo Cinelli, and Samuel Martin-Gutierrez for helpful feedback on the first draft of our manuscript. 

\end{document}